\documentclass[runningheads]{llncs}
\usepackage{graphicx}
\usepackage[inkscapelatex=false]{svg}
\usepackage{amsmath,amssymb,amsfonts}
\usepackage{algorithm}
\usepackage{algorithmic}
\usepackage{graphicx}
\usepackage{textcomp}
\usepackage{xcolor}
\usepackage[caption=false, font=footnotesize]{subfig}
\usepackage[title,toc,titletoc,header]{appendix}

\begin{document}
\title{Goal-Driven Context-Aware Next Service Recommendation for Mashup Composition}
\titlerunning{Goal-Driven Context-Aware Service Recommendation}
%
\author{}
\author{Xihao Xie\inst{1} \and
Jia Zhang\inst{1} \and
Rahul Ramachandran\inst{2} \and Tsengdar J. Lee\inst{3} \and Seungwon Lee\inst{4}}
\authorrunning{X. Xie et al.}
%

\institute{
Department of Computer Science, Southern Methodist University, USA \\ \email{\{xihaox, jiazhang\}@smu.edu}
\and
NASA/MSFC, USA\\ \email{rahul.ramachandran@nasa.gov}
\and
Science Mission Directorate, NASA Headquarters, USA\\ \email{tsengdar.j.lee@nasa.gov}
\and
NASA/JPL, USA\\ \email{seungwon.lee@jpl.nasa.gov}
}
\maketitle              
%
\begin{abstract}
As service-oriented architecture becoming one of the most prevalent techniques to rapidly deliver functionalities to customers, increasingly more reusable software components have been published online in forms of web services. To create a mashup, it gets not only time-consuming but also error-prone for developers to find suitable services from such a sea of services. Service discovery and recommendation has thus attracted significant momentum in both academia and industry. This paper proposes a novel incremental \emph{recommend-as-you-go} approach to recommending next potential service based on the context of a mashup under construction, considering services that have been selected to the current step as well as its mashup goal. The core technique is an algorithm of learning the embedding of services, which learns their past goal-driven context-aware decision making behaviors in addition to their semantic descriptions and co-occurrence history. A goal exclusionary negative sampling mechanism tailored for mashup development is also developed to improve training performance. Extensive experiments on a real-world dataset demonstrate the effectiveness of our approach.

\vspace{-6pt}
\end{abstract}
\vspace{-15pt}
\section{Introduction}
\vspace{-10pt}
In the last two decades, a huge number of software components have been published onto the Internet in forms of software services (or so-called APIs or services). Up to June 15th, 2022, ProgrammableWeb\footnote{https://www.programmableweb.com/}, the largest online repository of APIs, has accumulated 24,471 services. Such remotely accessible services enable software developers to compose existing services into mashups, easier and faster than before without creating everything from scratch. However, such a sea of services makes it time-consuming and error-prone for mashup developers to select suitable service candidates. Thus, service recommendation-powered mashup development has attracted significant momentum in recent years.

\begin{figure}[htbp]
\centerline{\includegraphics[width=\textwidth]{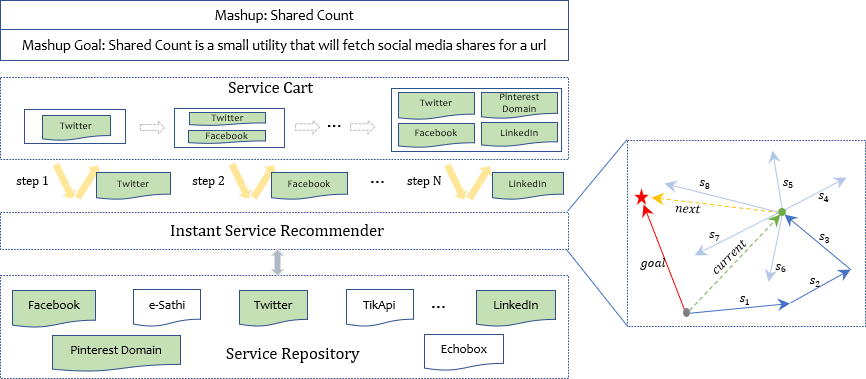}}
\caption{Incremental process of mashup creation.}
\label{fig_process}
\vspace{-15pt}
\end{figure}

Fig. \ref{fig_process} illustrates a possible development process for a mashup randomly selected from ProgrammableWeb named \emph{Shared Count}\footnote{https://www.programmableweb.com/mashup/shared-count}. As shown in Fig. \ref{fig_process}, a mashup creation is an incremental process with multiple rounds of service selection, e.g., \emph{Twitter} is selected first followed by \emph{Facebook}. 
In each round, an instant recommender suggests the ``next" suitable services. 

\vspace{-0pt}

In our position, an effective recommender shall guide a mashup development process as a journey of exploration in a forest (i.e., an embedding space), led by a light beam (i.e., the mashup goal), as illustrated in the right-hand side dotted box of Fig. \ref{fig_process}. At any point of decision making, e.g., the green dot, the services that have already been selected (i.e., \emph{Twitter} at step $t1$ and \emph{Facebook} at step $t2$) represent the position of the ``traveler" at the time. The problem thus becomes how to make the next step (i.e., select the next service) toward the ultimate goal (i.e., the red star representing the embedding of the textual mashup goal ``shared count is a small utility that will fetch social media shares for a url"). We coin a term \emph{context} to represent a decision making point during a mashup development process, which comprises selected services and the mashup goal each being denoted as a \emph{contextual item} throughout this paper.

To tackle this problem, this paper proposes a goal-driven, context-aware machine learning method capable of recommending ``next" suitable services in each step of the incremental process of mashup creation. In contrast to the state-of-the-art NLP-based word embedding \cite{b10} and transaction-based shopping cart embedding and recommendation \cite{b26}, we favor both mashup context and goal synergistically. Our approach mainly comprises two modules: an offline module of representation learning and an online module of next candidate service recommendation. In the offline module, we learn embeddings of contextual items and a global attention vector over a historical mashup repository. Note that each contextual service 
may make a different contribution scale to recommending next service, and we apply an attention mechanism to learn such scales. In the online module, once the context of an ongoing mashup is embedded based on the trained parameters, we conduct matrix calculation to rank the probabilities of potential services in descending order. Top K services will be recommended for the next round to help mashup developers speed up service selection.

To the best of our knowledge, we make the first effort to learning the embedding of goal-driven mashup context. For each service, in addition to learning its semantic descriptions and service co-occurrence, we learn its decision making behaviors from past mashup development provenance. Furthermore, our proposed learning framework is extensible for considering other dimensions of data such as user profile. The main contributions of our work are three-fold. First, we propose a novel machine learning algorithm that is capable of learning goal-driven context embedding during mashup development. Second, we propose a goal-exclusionary negative sampling strategy tailored for a rapid training process for mashup development. Third, we demonstrate the effectiveness of our approach based on extensive experiments on a real-world dataset.

The remainder of this paper is organized as follows. Section 2 summaries the related work. Section 3 defines the problem and presents our approach in detail. Section 4 explains how to train and learn relevant parameters. Section 5 presents and analyzes experimental results. Finally, Section 6 closes with conclusions.

\vspace{-2pt}
\section{Related Work}
\vspace{-8pt}
In this section, we will discuss closely related work from two aspects: service representation learning and next service recommendation.

\vspace{-5pt}
\subsection{Service Representation Learning}
\vspace{-5pt}

In recent years, researchers have explored many ways to represent services in a vector space to support downstream service recommendation applications. 
Here are some examples. Gu et al. \cite{b6} propose an approach to recommending a bundle of services by leveraging latent Dirichlet allocation (LDA) to embed services and mashups in a topic space. 
%
Wang et al. \cite{b37} propose Service2vec to learn service representations based on a constructed service network. Zhang et al. \cite{b38.1} incorporate service users’ perceptions into service profiles to form more comprehensive service representations. Menzi et al. \cite{b39} embed services into a low-dimensional vector space based on a constructed context-aware service knowledge graph to support service recommendation.

In contrast to their work, our approach learns service embeddings from past service selection decision making behaviors, in addition to service and mashup descriptions and correlations.

\vspace{-5pt}
\subsection{Next Service Recommendation}
\vspace{-6pt}
Next service recommendation aims to recommend candidate services for developers in each step of a multi-round process of mashup creation. Cao et al. \cite{b3} design a collaborative-filtering (CF)-based algorithm based on a two-level topic model to recommend services. Zhang et al. \cite{b20} propose an approach to extracting relationships of people, services and workflows from historical usage data into a social network to proactively recommend services in a workflow composition process. Liu et al. \cite{b21} develop a generalized sequential pattern algorithm to mine frequent composition patterns of mashups, and design an interactive recommendation algorithm to assist mashup creation. 
By modeling the relations between services and service-based systems (SBSs) into a heterogeneous information network, Xie et al. \cite{b24} measure semantic similarities between SBSs and use content filtering technique to recommend next service. Kirubananthan et al. \cite{b41} propose a method to support long-term service composition recommendation according to user ratings.

Our work differs from the aforementioned studies in two aspects. First, we introduce a model to embed the context including not only existing services but also mashup goals. In this way, our model can learn goal-driven next service prediction behaviors from past experiences. Second, we model different contributions of different contextual items with an attention mechanism.
\vspace{-8pt}
\section{Context-aware Next Service Recommendation}
\vspace{-8pt}

\vspace{-5pt}
\begin{figure*}[htbp]
\centerline{\includegraphics[width=\textwidth]{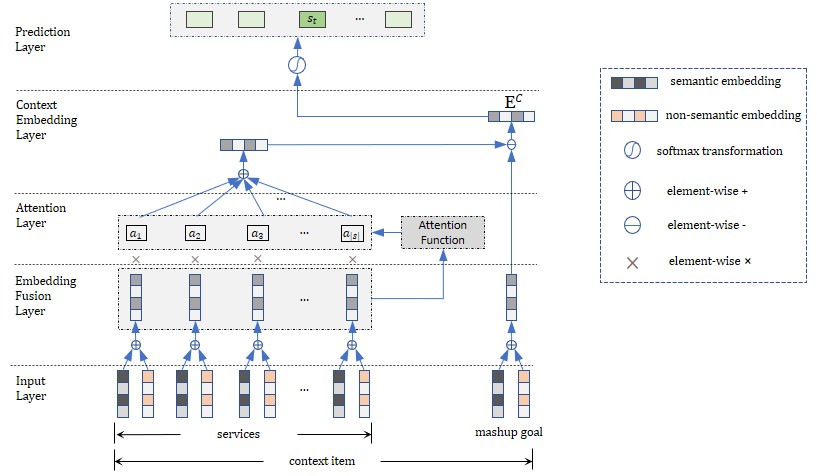}}
\vspace{-5pt}
\caption{Blueprint of proposed approach.
}
\vspace{-10pt}
\label{fig_overview}
\end{figure*}
\vspace{-2pt}

Fig.~\ref{fig_overview} depicts an overview of our goal-driven context-aware service recommendation framework. Given a partial mashup under construction, all services already selected and the mashup goal form the context of the recommendation problem.
As shown in Fig.~\ref{fig_overview}, for each contextual service item, its semantic embedding and auxiliary embedding (to be learned) are fused into an intermediate vector. Based on the services selected, an attention function (to be learned) will generate attention coefficients for them. All weighted service embeddings will be fused and, then fused with the intermediate vector from the semantic embedding and auxiliary embedding (to be learned) of the mashup goal. A fully connected layer with a weight matrix (to be learned) will compute the possibility of each service becoming the next service, and make the prediction via a softmax function.

Take the motivating mashup in Fig.~\ref{fig_process} again as an example, with $m$ being its textually described goal. At its second creation step, two services have been selected as contextual services, $s_{i}$ being \emph{Twitter} and $s_{k}$ being \emph{Facebook}. After training, the framework in Fig.~\ref{fig_overview} will recommend a candidate service,  $s_{n}$ being \emph{LinkedIn}, for the third step.


In this section, we first define the research problem and introduce some preliminary, and then discuss in detail our approach.

\vspace{-8pt}
\subsection{Research Problem Statement and Preliminary}
\vspace{-5pt}

Our study is based on a repository comprising all historical mashups and their component services. Let $\mathcal{M} = \left\{m_{1}, m_{2}, ..., m_{\left|\mathcal{M}\right|}\right\}$ be a set of mashups, each of which is textually described in natural language as its goal, and {$\mathcal{S} = \left\{s_{1}, s_{2}, ..., s_{\left|\mathcal{S}\right|}\right\}$} be a set of services. Each mashup $m_{i}$ contains a subset of services $S^{m_{i}} = \left\{s^{i}_{1}, s^{i}_{2}, ..., s^{i}_{\left|S^{m_{i}}\right|}\right\} \subseteq \mathcal{S}$. 
In an incremental process of creating mashup $m' \notin \mathcal{M}$, before it is completed, given multiple services that have been selected as contextual services ${S^{m'}_{t} = \left\{s^{m'}_{1}, s^{m'}_{2}, ..., s^{m'}_{t}\right\} \subseteq \mathcal{S}}$, where $s^{m'}_{j} \in \mathcal{S}, j \in \left[1, t\right]$ is the service selected at the $j\textsuperscript{th}$ step, our goal is to calculate the probability for each service $s \in \mathcal{S} \setminus S^{m'}_{t}$ and recommend top K candidate services at the next step $t + 1$. Further, let $C^{m'}_{t} = S^{m'}_{t} \cup \left\{m'\right\}$ denote the whole context at the $t\textsuperscript{th}$ step of creating mashup $m'$. In this way, the recommendation problem can be formulated to a problem of probabilistic classification by learning to predict a conditional probability:
\begin{equation}
{p(s^{m'}_{t + 1}=s | C^{m'}_{t}) = p(s | C^{m'}_{t})}
\label{eq_1}
\end{equation}

To solve the research problem, we first introduce some preliminary.

\vspace{-8pt}
\paragraph{Definition 1: Semantic Embedding Matrix.} Let matrix $\mathbf{E}^{S} \in \mathbb{R}^{(\left|\mathcal{S}\right| + 1) \times d}$ be a semantic vector space for services and mashup goal. The $j\textsuperscript{th}$ row of the matrix $\mathbf{E}^{S}_{j}$, $j \in \left[1, \left|\mathcal{S}\right|\right]$, is a d-dimension vector semantically representing the $j\textsuperscript{th}$ service $s_{j} \in \mathcal{S}$. The last row of $\mathbf{E}^{S}$, i.e., $\mathbf{E}^{S}_{|\mathcal{S}|+1}$, is the semantic embedding for mashup goal. This matrix aims to learn from textual descriptions of all services and mashups from the repository.

\vspace{-8pt}
\paragraph{Definition 2: Auxiliary Embedding Matrix.} Let matrix $\mathbf{E}^{X} \in \mathbb{R}^{(\left|\mathcal{S}\right| + 1) \times d}$ denote an auxiliary vector space for services and mashup goal. The $j\textsuperscript{th}$ row of the matrix $\mathbf{E}^{X}_{j}$, $j \in \left[1, \left|\mathcal{S}\right|\right]$, is a d-dimension vector representation for the $j\textsuperscript{th}$ service $s_{j} \in \mathcal{S}$. The last row of $\mathbf{E}^{X}$, i.e., $\mathbf{E}^{X}_{|\mathcal{S}|+1}$ is the auxiliary embedding for mashup goal. This matrix aims to learn from past service occurrence and decision making behaviors.

\vspace{-8pt}
\paragraph{Definition 3: Attention Vector.} Let $\mathbf{A} \in \mathbb{R}^{d}$ be a global d-dimension vector that is shared by all services. It is used to model the contribution scales of contextual services based on an attention mechanism.

\vspace{-8pt}
\paragraph{Definition 4: Weight Matrix.} Let $\mathbf{E}^{W} \in \mathbb{R}^{d \times \left|\mathcal{S}\right|}$ be a weight matrix of the fully connected layer, as shown in Fig. ~\ref{fig_overview}. It is used to help predict the conditional probability defined in Eq.~\eqref{eq_1}.

In our framework, $\Theta = \{\mathbf{E}^{X}$, $\mathbf{A}$, $\mathbf{E}^{W}\}$ will be learned through back-propagation, which will be discussed in detail in the following sections.

\vspace{-5pt}
\subsection{Context Embedding}
\vspace{-5pt}
The semantic embedding matrix $\mathbf{E}^{S}$ is learned through NLP machine learning methods such as the Doc2Vec \cite{b34} model. Each textual description of mashup goal or service is regarded as a document. The corpus of all documents is fed into the Doc2Vec model to receive a vectorized representation of each mashup goal or service. Specially, we adopt the distributed memory model of paragraph vectors (PV-DM) as it performs better than the distributed bag of words of paragraph vector (PV-DBOW) \cite{b34}. The trained $\mathbf{E}^{S}$ will also be used for goal exclusionary negative sampling, which will be discussed later in Section 4.

The auxiliary embedding matrix $\mathbf{E}^{X}$ carries non-semantic information, e.g., co-occurrence information and decision-making strategies, of services and mashup. For any two services items for example, the closer they are in $\mathbf{E}^{X}$, the more likely they are to co-occur in the same context. Note that $\mathbf{E}^{X}_{|\mathcal{S}| + 1}$, the last row of the matrix, is an in-situ vector that will be shared by all mashups. The vector will be trained as an offset vector to adjust the semantic representation of a mashup goal in a decision making process.

Given a partial mashup $C^{m'}_{t}$, the input units in the input layer of Fig.~\ref{fig_overview} constitute a one-hot encoding vector where only the unit at position $i$ ($s_{i} \in C^{m'}_{t}$) is set to 1. Thus, for each contextual service $s_{i} \in C^{m'}_{t}$, the corresponding $i\textsuperscript{th}$ rows of the trained $\mathbf{E}^{S}$ and $\mathbf{E}^{X}$, $i \in [1, |\mathcal{S}|]$, are fed into the fusion layer. For the mashup goal, its semantic embedding $\mathbf{E}^{S}_{|\mathcal{S}| + 1}$ will be obtained in an \emph{ad hoc} manner from the trained Doc2Vec model, and its auxiliary embedding will be obtained from the trained $\mathbf{E}^{X}_{|\mathcal{S}| + 1}$. 

As shown in Fig.~\ref{fig_overview}, for each $i\textsuperscript{th}$ contextual item, its semantic embedding and auxiliary embedding are fused into a fused vector in the fusion layer: 
$\mathbf{E}^{F}_{i} = f(\mathbf{E}^{S}_{i}, \mathbf{E}^{X}_{i})$, where $i \in \left[1, \left|\mathcal{S}\right| + 1\right]$, and $f: \mathbb{R}^{d} \times \mathbb{R}^{d} \mapsto \mathbb{R}^{d}$ is a fusion function.

\vspace{-5pt}
\subsection{Attention-based Context Embedding}
\vspace{-8pt}

Considering that different contextual services may weight differently in representing the whole context, 
we model the contribution scale of every contextual service with an attention mechanism. Specifically, for each contextual service $s^{m'}_{t, k} \in C^{m'}_{t}$, its contribution scale is denoted as the weight of $w^{m'}_{t, k}$, where $k \in \left[1, \left|\mathcal{S}\right|\right]$. It can be calculated with a softmax transformation which is widely used in neural networks:

\vspace{-8pt}
\begin{equation}
w^{m'}_{t, k} = \frac{\exp(\Lambda(k))}{\sum_{s^{m'}_{t, h} \in C^{m'}_{t}}\exp(\Lambda(h))}
\label{eq_2}
\end{equation}
\vspace{-3pt}
\begin{equation}
\Lambda(k) = \mathbf{E}^{F}_{k} \cdot \mathbf{A} = f\left(\mathbf{E}^{S}_{k}, \mathbf{E}^{X}_{k}\right) \cdot \mathbf{A} 
\label{eq_3}
\end{equation}

where $\Lambda(k)$ is an attention function shown in Fig.~\ref{fig_overview}. It is a dot product of the fused vector $\mathbf{E}^{F}_{k}$ with an attention vector $\mathbf{A}$ to be learned, which is used to model the contribution scales of contextual services.

The resulted integration weight $w^{m'}_{t, k}$ acts as the attention scale of contextual service $s^{m'}_{t, k}$. The higher $w^{m'}_{t, k}$ is, the more we should pay attention to the contextual service $s^{m'}_{t, k}$ than other services $s^{m'}_{t, l} \in C^{m'}_{t} \setminus \{s^{m'}_{t, k}\}$ at the $t\textsuperscript{th}$ step of creating mashup $m'$. 

As shown in Fig. \ref{fig_overview}, we integrate the representations of all contextual services and that of mashup goal to represent the whole context $C^{m'}_{t}$ as follows:

\vspace{-10pt}
\begin{equation}
\mathbf{E}^{C^{m'}_{t}} = \mathbf{E}^{F}_{|\mathcal{S}| + 1} + \sum_{s^{m'}_{t, k} \in C^{m'}_{t}} w^{m'}_{t, k} \times \mathbf{E}^{F}_{k}, \quad s.t. \sum_{s^{m'}_{t, k} \in C^{m'}_{t}} w^{m'}_{t, k} = 1
\label{eq_4}
\end{equation}
\vspace{-5pt}

\vspace{-12pt}
\subsection{Prediction \& Recommendation}
\vspace{-5pt}

Since the context $C^{m'}_{t}$ has been 
represented as Eq. \ref{eq_4}, the conditional probability of Eq. \ref{eq_1} becomes:
\begin{equation}
p(s^{m'}_{t + 1} = s | C^{m'}_{t}) = p(s | \mathbf{E}^{C^{m'}_{t}})
\label{eq_5}
\end{equation}

As shown in Fig. \ref{fig_overview}, in the fully connected layer, the $j\textsuperscript{th}$ column of the weight matrix (to be learned) $\mathbf{E}^{W}_{j}$ is used to calculate a scalar value to help obtain the conditional probability. Specifically, for every service $s_{n} \in \mathcal{S}$, the scalar value can be calculated as a dot production of the context embedding $\mathbf{E}^{C^{m'}_{t}}$ and $\mathbf{E}^{W}_{n}$:
\begin{equation}
v(n) = \mathbf{E}^{C^{m'}_{t}} \cdot \mathbf{E}^{W}_{n}
\label{eq_6}
\end{equation}

Here $v(n)$ can be seen as a scoring function qualifying the relevance of a service $s_{n}$ with respect to the given context $C^{m'}_{t}$. For a service $s_{n}$, the higher the score is, the more relevant it is to the context, and the more likely it should be recommended. Thereafter, for any service $s_{n} \in \mathcal{S}$, the likelihood of it to be selected at the next step defined in Eq. \ref{eq_5} can be further normalized as a result of the softmax transformation:
\begin{equation}
p_{\Theta}(s^{m'}_{t+1} = s_{n} | \mathbf{E}^{C^{m'}_{t}}) = \frac{\exp(v(n))}{\sum_{s^{m'}_{t, h} \in \mathcal{S}}\exp(v(h))}
\label{eq_7}
\end{equation}

Now that the probability of a service to be selected at the next step has been calculated, for all of the services in $\mathcal{S} \setminus S^{m'}_{t}$ that have not been selected, we rank them according to their probabilities in descending order, and recommend the top K services. 

In practice, at the beginning of creating a mashup, only the mashup goal will be used as the input of our model. 
That is, according to Eq.~\eqref{eq_4}, the embedding of context becomes: $\mathbf{E}^{C^{m'}_{0}} = \mathbf{E}^{F}_{|\mathcal{S}| + 1} = f(\mathbf{E}^{S}_{|\mathcal{S}| + 1}, \mathbf{E}^{X}_{|\mathcal{S}| + 1})$.

\vspace{-0pt}
\section{Model Learning}
\vspace{-10pt}
In this section, we explain how to train our model, i.e., $\Theta=\{\mathbf{E}^{X}$, $\mathbf{A}, \mathbf{E}^{W}\}$. The matrices are learned throughout creation steps of mashups from historical data offline.



\vspace{-10pt}
\subsection{Objective Function}
\vspace{-5pt}
Let $\mathcal{D} = \left\{D^{i}\right\}$ denote a training data set, where $D^{i} = \bigcup\limits_{0 \leq t \textless |S^{m_{i}}|} \{D^{i}_{t}\}$; $D^{i}_{t} = \left<C^{i}_{t}, G^{i}_{t}\right>$ is a training instance from mashup $m_{i}$ at the $t\textsuperscript{th}$ step of creating it; $C^{i}_{t} = \left\{m_{i}\right\} \cup S^{m_{i}}_{t}$ is the given context, and $G^{i}_{t} \subseteq \mathcal{S} \setminus S^{m_{i}}_{t}$ is the set of the observed ground truth services that co-occur with $S^{m_{i}}_{t}$ in the mashup at step $t$. Given a training set $\mathcal{D}$, the joint probability distribution can be obtained as:
\vspace{-5pt}
\begin{equation}
    p_{\Theta}(\mathcal{D}) \propto \prod_{D^{i} \in \mathcal{D}}  \prod_{D^{i}_{t} \in D^{i}}p_{\Theta}(G^{i}_{t} | \mathbf{E}^{C^{i}_{t}})
\end{equation}
\vspace{-8pt}

Thereafter, parameters $\Theta$ can be learned by maximizing the following objective function over all training instances:
\vspace{-8pt}
\begin{small}
\begin{equation}
L_{\Theta} = \log\prod_{D^{i} \in \mathcal{D}}\prod_{D^{i}_{t} \in D^{i}}\sum_{s^{i}_{t,j} \in G^{i}_{t}}p_{\Theta}(s^{i}_{t,j} | \mathbf{E}^{C^{i}_{t}}) = \log\prod_{D^{i} \in \mathcal{D}}\prod_{D^{i}_{t} \in D^{i}}\frac{\sum_{s^{i}_{t,j} \in G^{i}_{t}}\exp(v(j))}{\sum_{s_{h} \in \mathcal{S}}\exp(v(h))}
\label{eq_8}
\end{equation}
\end{small}
\vspace{-8pt}

where $p_{\Theta}(s^{i}_{t,j} | \mathbf{E}^{C^{i}_{t}})$ is the conditional probability according to Eq. \ref{eq_7}.

\vspace{-2pt}
\subsection{Sampling Strategy}
\vspace{-5pt}
Optimizing the objective function in Eq. \ref{eq_8}, however, is non-trivial since each evaluation of the softmax function has to traverse all services, which is extremely time-consuming. To learn the aforementioned parameters efficiently, we employ the idea of negative sampling \cite{b25} to approximate the conditional probability. Considering that a service being negative to a mashup $m_{i}$ tends to be less compatible to $m_{i}$, we introduce a goal exclusionary negative sampling strategy to obtain a set of negative services. Specifically, based on the trained Doc2Vec model, for all services in $\mathcal{S} \setminus S^{m_{i}}$, we rank them according to their cosine-based semantic similarities with the mashup goal in ascending order and choose the top $r$ percent as negative samples. In this way, let $N^{i} \subseteq \mathcal{S} \setminus S^{m_{i}}$ denote the set of negatively sampled services, each of which is not contained in the mashup $m_{i}$. Then, we redefine the conditional probability as:
\vspace{-5pt}
\begin{equation}
p_{\Theta}(G^{i}_{t}, N^{i} | \mathbf{E}^{C^{i}_{t}}) = \prod_{s_{h} \in G^{i}_{t} \cup N^{i}} p_{\Theta}(s_{h} | \mathbf{E}^{C^{i}_{t}})
\label{eq_10}
\end{equation}
\vspace{-8pt}

\vspace{-5pt}
\begin{equation}
p_{\Theta}(s_{h} | \mathbf{E}^{C^{i}_{t}}) =
\begin{cases}
\sigma(v_{\Theta}(h)) & s_{h} \in G^{i}_{t} \\
1 - \sigma(v_{\Theta}(h)) & s_{h} \in N^{i}
\end{cases}
\label{eq_11}
\end{equation}
\vspace{-2pt}
where $\sigma$ is the sigmoid function $\sigma(x) = 1 / (1 + \exp(-x))$ and $v_{\Theta}(h)$ is the scoring function following the definition of Eq. \ref{eq_6}. In Eq. \ref{eq_11}, $\sigma(v_{\Theta}(h))$ can be seen as the probability of service $s_{h}$ being labeled as a ground truth service with a score of $v_{\Theta}(h)$ given context of $C^{i}_{t}$. Thus, the probability of it being labeled as a negative service is $1 - \sigma(v_{\Theta}(h))$. In this way, maximizing the conditional probability in Eq. \ref{eq_10} means maximizing the probability of a service being a positive sample and in the meantime minimizing the probability of it being a negative sample. After that, the objective function becomes:

\vspace{-5pt}
\begin{small}
\begin{equation}
\begin{aligned}
L_{\Theta} &= \log\prod_{D^{i} \in \mathcal{D}}\prod_{D^{i}_{t} \in D^{i}}\prod_{s_{h} \in G^{i}_{t} \cup N^{i}}p_{\Theta}(s_{h} | \mathbf{E}^{C^{i}_{t}}) \\
&= \sum_{D^{i} \in \mathcal{D}} \sum_{D^{i}_{t} \in D^{i}} \left\{\sum_{s_{h} \in G^{i}_{t}}\log\sigma(v_{\Theta}(h)) + \sum_{s_{h} \in N^{i}}\log[\sigma(-v_{\Theta}(h))]\right\}
\end{aligned}
\label{eq_12}
\end{equation}
\end{small}

At last, according to Eq. \ref{eq_12}, we utilize the stochastic gradient ascent to learn the aforementioned parameters through back-propagation. Let $L^{i,t}_{\Theta}$ denote the objective function for a training instance $D^{i}_{t}$ as:

\vspace{-5pt}
\begin{equation}
    L^{i,t}_{\Theta} = \sum_{s_{h} \in G^{i}_{t}}\log\sigma(v_{\Theta}(h)) + \sum_{s_{h} \in N^{i}}\log[\sigma(-v_{\Theta}(h))]
    \label{eq_13}
\end{equation}
\vspace{-5pt}

The gradient and update rule of $\mathbf{E}^{W}_{h}$ then can be represented as:
\vspace{-8pt}
\begin{equation}
    \frac{\partial L^{i,t}_{\Theta}}{\partial \mathbf{E}^{W}_{h}} = [l(h) - \sigma(v_{\Theta}(h))] \cdot \mathbf{E}^{C^{i}_{t}}, \mathbf{E}^{W}_{h} \gets \mathbf{E}^{W}_{h} + \eta \cdot \frac{\partial L^{i,t}_{\Theta}}{\partial \mathbf{E}^{W}_{h}}
    \label{eq_14}
\end{equation}
\vspace{-5pt}

where $l(h)$ is a label function whose value is 1 when $s_{h} \in G^{i}_{t}$ or 0 otherwise. The gradient and update rule of $\mathbf{A}$ for $D^{i}_{t}$ can be represented as:

\vspace{-8pt}
\begin{equation}
    \frac{\partial L^{i,t}_{\Theta}}{\partial \mathbf{A}} = [l(h) - \sigma(v_{\Theta}(h))] \cdot \mathbf{E}^{W}_{h} \cdot \frac{\partial \mathbf{E}^{C^{i}_{t}}}{\partial \mathbf{A}}, \mathbf{A} \gets \mathbf{A} + \eta \cdot \frac{\partial L^{i,t}_{\Theta}}{\partial \mathbf{A}}
    \label{eq_15}
\end{equation}
\vspace{-8pt}

Similarly, the gradient and update rule of $\mathbf{E}^{X}_{k}$ for $s_{k} \in S^{m_{i}}_{t}$ and $k = |\mathcal{S}| + 1$ becomes:
\vspace{-8pt}
\begin{equation}
    \frac{\partial L^{i,t}_{\Theta}}{\partial \mathbf{E}^{X}_{k}} = [l(h) - \sigma(v_{\Theta}(h))] \cdot \mathbf{E}^{W}_{h} \cdot \frac{\partial \mathbf{E}^{C^{i}_{t}}}{\partial \mathbf{E}^{X}_{k}}, \mathbf{E}^{X}_{k} \gets \mathbf{E}^{X}_{k} + \eta \cdot \frac{\partial L^{i,t}_{\Theta}}{\partial \mathbf{E}^{X}_{k}}
    \label{eq_16}
\end{equation}
\vspace{-10pt}
For the auxiliary embedding of mashup goal, $\frac{\partial \mathbf{E}^{C^{i}_{t}}}{\partial \mathbf{E}^{X}_{|\mathcal{S}| + 1}} = 1 \cdot \frac{\partial\mathbf{E}^{F}_{|\mathcal{S}| + 1}}{\partial\mathbf{E}^{X}_{|\mathcal{S}| + 1}}$. 

\vspace{-0pt}
\subsection{Model Training} \label{sec_model_training}
Algorithm 1 lists the pseudo code of model training. The first two steps are preprocessing: we first initialize parameters ($\mathbf{E}^{X}$, $\mathbf{A}$ and $\mathbf{E}^{W}$) randomly; and then feed the corpus of textual descriptions of mashup goals and services into the PV-DM model \cite{b34} to obtain semantic embedding of services, i.e., $\mathbf{E}^{S}_{:|\mathcal{S}|}$. Afterwards, for each mashup $m_{i}$ in the training set (line 5), we obtain its semantic embedding (line 6), i.e., $\mathbf{E}^{S}_{|\mathcal{S}| + 1}$, from the PV-DM model, and sample the top $r$ percent goal exclusionary negative services (line 7). After that, incrementally, at every creation step $t$ (line 8), we generate a training instance $D^{i}_{t}$ (line 9), containing contextual services $S^{m_{i}}_{t}$, mashup goal, and ground truth services $G^{i}_{t}$. Specifically, $S^{m_{i}}_{t}$ is augmented by randomly choosing one service from $G^{i}_{t - 1}$. Subsequently, $G^{i}_{t}$ becomes $S^{m_{i}} \setminus S^{m_{i}}_{t}$. Afterwards, we embed the context as a d-dimension vector according to Eq. \eqref{eq_4} (line 10). We then compute the objective function value according to Eq. \eqref{eq_13} (line 11), and update $\mathbf{E}^{X}$, $\mathbf{A}$ and $\mathbf{E}^{W}$ (line 12) according to Eqs. \eqref{eq_14}, \eqref{eq_15} and \eqref{eq_16} respectively. Finally, we stop learning if the accumulated objective function values converge.

\begin{algorithm}[H]
\caption{Learning parameters $\mathbf{E}^{X}$, $\mathbf{A}$ and $\mathbf{E}^{W}$ offline}
\begin{algorithmic}[1]
\renewcommand{\algorithmicrequire}{\textbf{Input:}}
\renewcommand{\algorithmicensure}{\textbf{Output:}}
\REQUIRE mashups $\mathcal{M}$, services $\mathcal{S}$ in training set, dimension size $d$, learning rate $\eta$ and negative sampling ratio $r$
\ENSURE parameters $\mathbf{E}^{X}$, $\mathbf{A} $ and $\mathbf{E}^{W}$
\STATE Initialize $\mathbf{E}^{X} \in \mathbb{R}^{(\left|\mathcal{S}\right| + 1) \times d}$, $\mathbf{A} \in \mathbb{R}^{d}$ and $\mathbf{E}^{W} \in \mathbb{R}^{d \times \left|\mathcal{S}\right|}$
\STATE $\mathbf{E}^{S}_{:|\mathcal{S}|} \gets PV-DM(\mathcal{M}, \mathcal{S}, d)//$ \small{obtain semantic embedding of services from PV-DM}
\STATE $L \gets 0$
\WHILE{$L$ not converged}
  \FOR{each mashup $m_{i} \in \mathcal{M}$}
    \STATE $\mathbf{E}^{S}_{|\mathcal{S}| + 1} \gets PV-DM(m_{i})$\\
    $//$ \small{obtain semantic embedding of mashup goal from PV-DM}
    \STATE $N^{i} \gets NegativeSampling(\mathbf{E}^{S}, r)$ $//$ \small{Sample top $r$ percent negative services}
    \FOR{each creation step $t$ of $m_{i}$}
      \STATE $D^{i}_{t} \gets GenerateInstances(m_{i}, \mathcal{S})$
       $//$ \small{generate $D^{i}_{t}$ as $\left<C^{i}_{t}, G^{i}_{t}\right>$}
      \STATE $\mathbf{E}^{C^{i}_{t}} \gets EmbedContext(\mathbf{E}^{S}, \mathbf{E}^{X}, \mathbf{A}, C^{i}_{t})//$ \small{context embedding using Eq. \eqref{eq_4}}
      \STATE $L^{i,t} \gets ObjectiveFunctionValue(\mathbf{E}^{W}, \mathbf{E}^{C^{i}_{t}},  G^{i}_{t}, N^{i})$\\
      $//$ \small{Compute value of objective function using Eq. \eqref{eq_13}}
      \STATE $UpdateParameters(G^{i}_{t}, N^{i}, \mathbf{E}^{C^{i}_{t}}, \eta)$ \\
      $//$ \small{Update parameters according to the rules in Eq. \eqref{eq_14}, \eqref{eq_15} and \eqref{eq_16}}
      \STATE $L \gets L + L^{i,t}$
    \ENDFOR
  \ENDFOR
\ENDWHILE
\RETURN $\mathbf{E}^{X}$, $\mathbf{A}$, $\mathbf{E}^{W}$
\label{alg_offline_learning}
\end{algorithmic}
\end{algorithm}
\vspace{-20pt}

Note that our offline training stage is goal driven in two aspects. First, the auxiliary embedding of mashup goal and contextual services, i.e., $\mathbf{E}^{X}_{|\mathcal{S}| + 1}$ and $\mathbf{E}^{X}_{:|\mathcal{S}|}$, are jointly and reciprocally learned based on the semantic embedding of mashup goal, i.e., $\mathbf{E}^{S}_{|\mathcal{S}| + 1}$. Second, in the negative sampling phase, mashup goal is leveraged to sample negative services that are semantically exclusionary to it.

\section{Experiments}
In this section, we explain experiments conducted and analyze the results.

\subsection{Experimental Setup}
\vspace{-8pt}
We designed three research questions to evaluate our approach from three angles:
\begin{itemize}
\item \textbf{RQ1}: How does our approach perform in terms of recommendation metrics?
\item \textbf{RQ2}: How does our approach perform throughout the incremental process?
\item \textbf{RQ3}: How does the attention mechanism affect the performance?
\end{itemize}



We evaluated our approach on a real-world dataset that was crawled from ProgrammableWeb up to December 2020. 
The mashups containing only one service were removed. Consequently, we obtained 1,553 mashups and 663 services to form our testbed. 
Table~\ref{tab_data_set} summarizes the overall statistics of our dataset.

\setlength{\tabcolsep}{10mm}
\begin{table}[htbp]
\vspace{-10pt}
\caption{Statistic Information of the Dataset}
\vspace{-15pt}
\label{tab_data_set}
\begin{center}
\begin{tabular}{l r}
\hline
Statistics & Values
\\\hline
\# of Mashups & 1,553
\\
\# of Services & 663
\\
Avg. \# of services per mashup & 2.70
\\
\# of Training instances & 3,310
\\
\# of Testing instances & 883
\\\hline
\end{tabular}
\end{center}
\end{table}


During the training stage, we randomly chose 80\% of mashups as the training set and the rest 20\% ones as the test set. 
Without losing generality, we built training instances at the last step of mashup construction. For each mashup $m_{i}$ in the training set, we constructed $\left|S^{m_{i}}\right|$ training instances. In detail, for mashup $m_{i}$, each service $s^{i}_{k} \in S^{m_{i}}$ was used as ground truth to be recommended, and the rest services of $S^{m_{i}}$ were treated as the corresponding contextual services. That is to say, $D^{i}$ was expanded as $D^{i} = \cup_{s^{i}_{k} \in m_{i}}\{D^{i}_{k}\}$, where $D^{i}_{k} = \left<C^{i}(k), s^{i}_{k}, N^{i}\right>$ and $C^{i}(k) = \{m_{i}\} \cup S^{m_{i}} \setminus \{s^{i}_{k}\}$ is the corresponding context of $s^{i}_{k}$.

For textual descriptions of services and mashup goals, we performed a series of pre-processing operations including tokenization, stop-words removal and lemmatization with the NLTK\footnote{http://www.nltk.org/} library. After that, we utilized the PV-DM model from the doc2vec library of Gensim\footnote{https://radimrehurek.com/gensim/models/doc2vec.html} to obtain semantic vectorized embeddings (i.e., $\mathbf{E}^{S}$). As for $N^{i}$, we set the negative sampling ratio (i.e., $r$) to be 20\%.

We adopted the \emph{leave-last-out} evaluation scheme in the experiments, meaning that we evaluated recommendation performance at the last step of constructing the testing mashups. Similar to the expansion operation adopted for the training set, for each mashup $m_{j}$ in the testing set, we expanded it to $\left|S^{m_{j}}\right|$ testing instances. For each $s^{j}_{k} \in S^{m_{j}}$, it was used as the ground truth service and the remainder of $S^{m_{j}}$, $C^{j}(k) = \{m_{j}\} \cup S^{m_{j}} \setminus \{s^{j}_{k}\}$, was used as corresponding context. 


Four competitive methods were used as the baselines:
\vspace{-3pt}
\begin{itemize}
\item \textbf{MatchUp} \cite{b30}: A service recommender system which is based on correlations of used services as the user context.
\item \textbf{SNRec} \cite{b31}: An approach to recommending next service according service similarities based on co-existence.
\item \textbf{WVSM} \cite{b32}: A content-based method to recommend services using textual content similarity.
\item \textbf{ISRec} \cite{b24}: A clustering and classification-based service recommendation approach with word embeddings.
\end{itemize}
\vspace{-5pt}

Note that, in our test bed, services that are published after the mashup were not available to be recommended. Thus, for all methods, the services whose submission time is earlier than that of the mashup were hold out for recommendation. In addition, for simplicity, we used element-wise plus as the fusion function: $\mathbf{E}^{F} = f(\mathbf{E}^{S}, \mathbf{E}^{X}) = \mathbf{E}^{S} \oplus \mathbf{E}^{X}$. Thus, in Eq.~\eqref{eq_16}, $\frac{\partial \mathbf{E}^{C^{i}_{t}}}{\partial \mathbf{E}^{X}_{|\mathcal{S}| + 1}} = 1$.

\vspace{-15pt}
\subsection{Evaluation Metrics}
\vspace{-10pt}

We employed the following two commonly used accuracy metrics for evaluation. For both metrics, the higher the value is, the better the performance is.
\vspace{-8pt}
\begin{itemize}
\item[1.] \textbf{REC@K}: It measures the recall of the top K recommended services over all testing set. We reported the results with $K \in \left\{3, 5, 10, 20\right\}$.
\begin{small}
\vspace{-8pt}
\begin{equation}
REC@K = \frac{\left|R^{K} \cap G\right|}{\left|G\right|}
\label{eq_17}
\vspace{-10pt}
\end{equation}
\end{small}
where 
$R^{K}$ and $G$ are the recommended list of services and the ground truth services, 
respectively.

\item[2.] \textbf{MRR}: It measures the reciprocal rank of the position of the ground truth over all testing set.
\vspace{-5pt}
\begin{small}
\begin{equation}
MRR = \frac{1}{rank(R^{K}, G)}
\label{eq_18}
\end{equation}
\end{small}
where 
$rank(R^{K}, G)$ is the hit position of the ground truth in the recommended result. 
\end{itemize}

\vspace{-18pt}
\subsection{Performance Comparison}
\vspace{-8pt}

For MatchUp, we defined the importance of a service as the number of mashups in which it is invoked \cite{b30}. The importance of a mashup was defined as the mean importance of its component services. For WVSM, being the same with \cite{b32}, we set the smooth efficient $\lambda$ to 0.1. For ISRec, except for the number of topics $T$, which was set to be 28, we set other parameters 
the same with \cite{b24}. For our approach, the dimension size $d$ and learning rate $\eta$ were set to be 100 and 0.001, respectively. Table \ref{tab_performance_comparison} shows the results of mean metric values of different methods over all testing instances. The best is in bold and the second best is underlined. In terms of all accuracy metrics, the experiment demonstrates that our approach outperforms others, which answers the first research question \textbf{RQ1}.

\setlength{\tabcolsep}{2mm}
\begin{table}[htbp]
\vspace{-15pt}
\caption{Performance comparison of different approaches}
\vspace{-15pt}
\label{tab_performance_comparison}
\begin{center}
\begin{tabular}{l c c c c c}
\hline
Methods & REC@3 & REC@5 & REC@10 & REC@20 & MRR
\\\hline
MatchUp & 0.2514 & 0.2899 & 0.3386 & 0.3613 & 0.2099
\\
SNRec & 0.2661 & 0.3205 & 0.3692 & 0.4145 & 0.1738
\\
WVSM & 0.2276 & 0.2820 & 0.3533 & 0.4723 & 0.1790
\\
ISRec & \underline{0.3228} & \underline{0.4009} & \underline{0.4881} & \underline{0.5946} & \underline{0.2325}
\\\hline
Ours & \textbf{0.3579} & \textbf{0.4602} & \textbf{0.5277} & \textbf{0.6116} & \textbf{0.2872}
\\\hline
\end{tabular}
\end{center}
\vspace{-20pt}
\end{table}
\vspace{-5pt}

Both MatchUp and SNRec take into account invocations between mashups and services. 
However, they do not consider mashup goal and the contribution scales of contextual services. 
The reasons that our approach performs better than WVSM 
are two-fold. First, our approach takes into account not only textual semantic information, but also decision-making information. Second, we take contributions of different contextual services into consideration. As for ISRec, our approach performs better because it does not make use of the different contribution scales of selected services. 

Especially, in terms of MRR, the results of other approaches are lower than 0.25. 
However, our approach reaches at 0.2872, meaning that generally the hit service will be ranked at the third or forth position of the recommended result, which is quite useful to support mashup development.

\vspace{-8pt}
\subsection{Recommendation Performance in Mashup Creation Process}
\vspace{-8pt}
\begin{figure*}[htbp]
\centering
\includegraphics[width=0.24\textwidth]{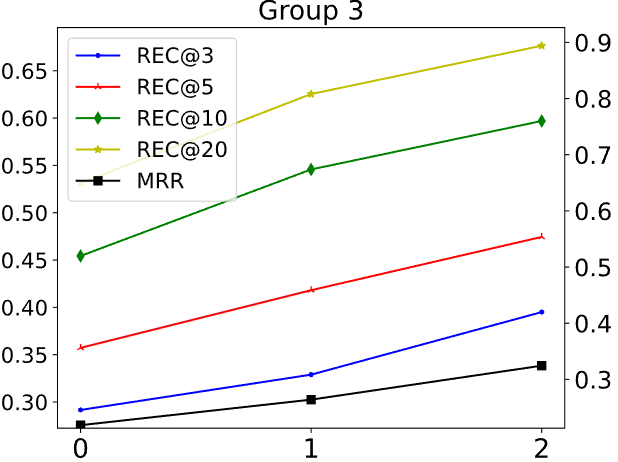}
\includegraphics[width=0.24\textwidth]{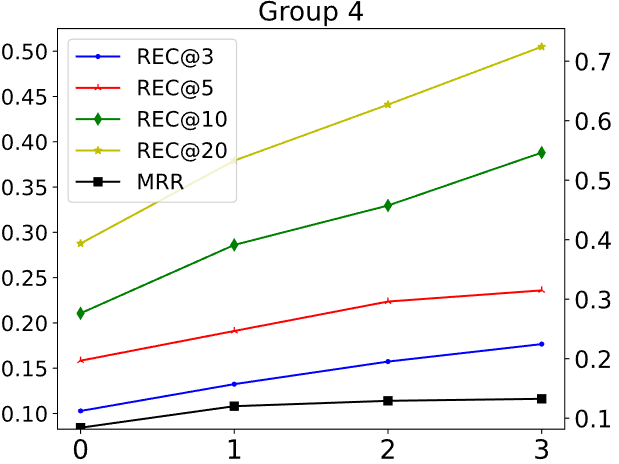}
\includegraphics[width=0.24\textwidth]{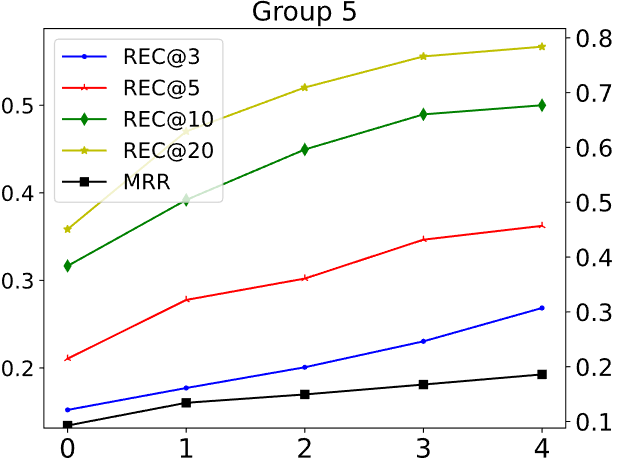}
\includegraphics[width=0.24\textwidth]{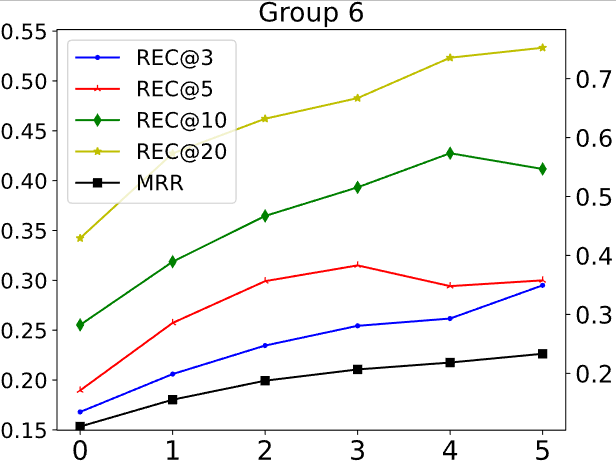}
\quad
\includegraphics[width=0.24\textwidth]{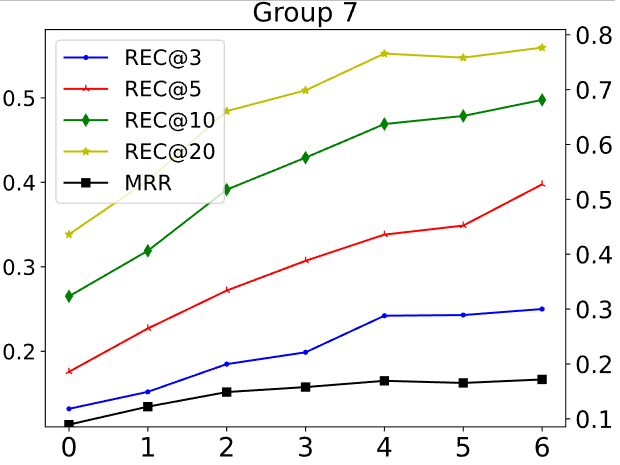}
\includegraphics[width=0.24\textwidth]{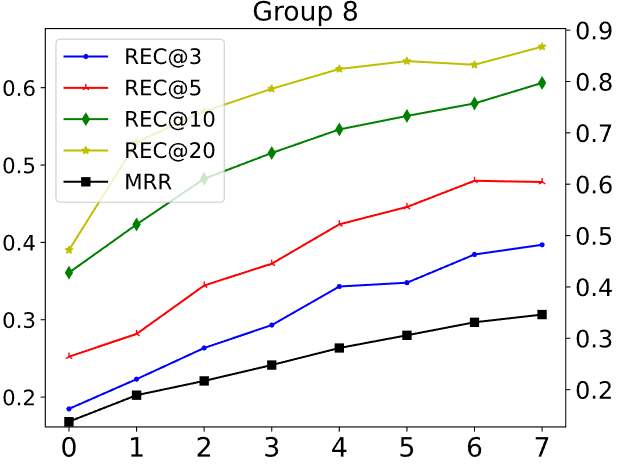}
\includegraphics[width=0.24\textwidth]{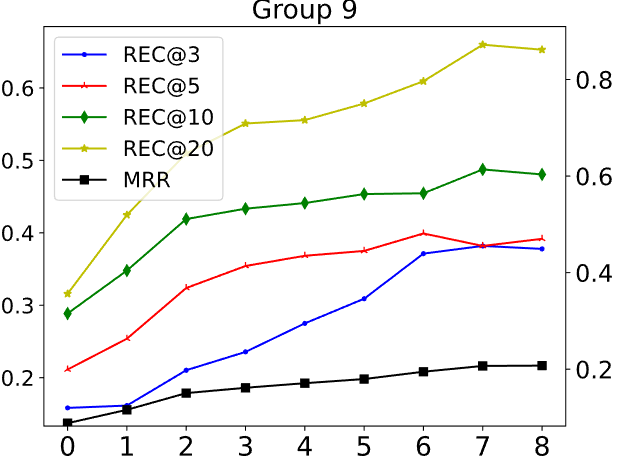}
\includegraphics[width=0.24\textwidth]{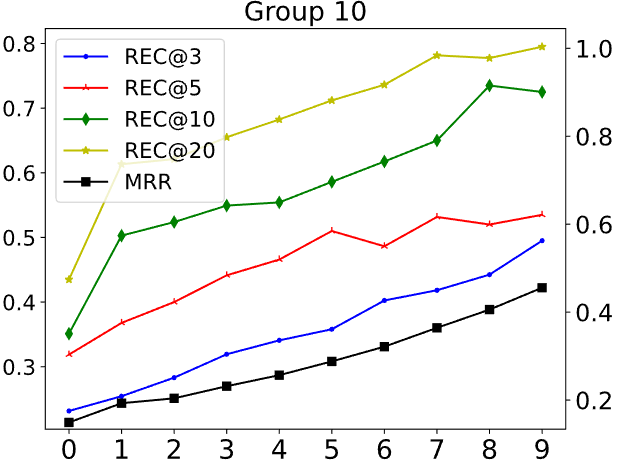}
\quad
\includegraphics[width=0.24\textwidth]{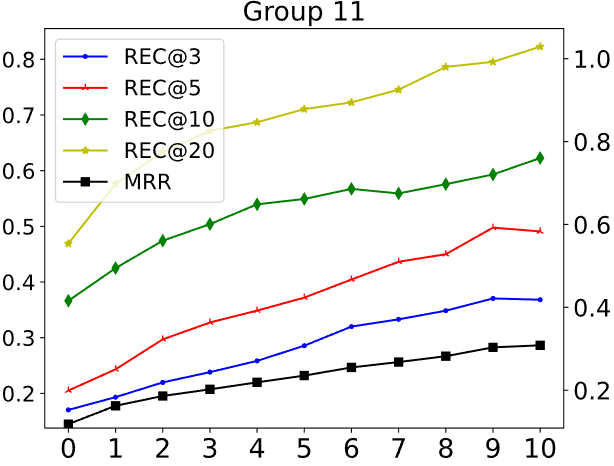}
\includegraphics[width=0.24\textwidth]{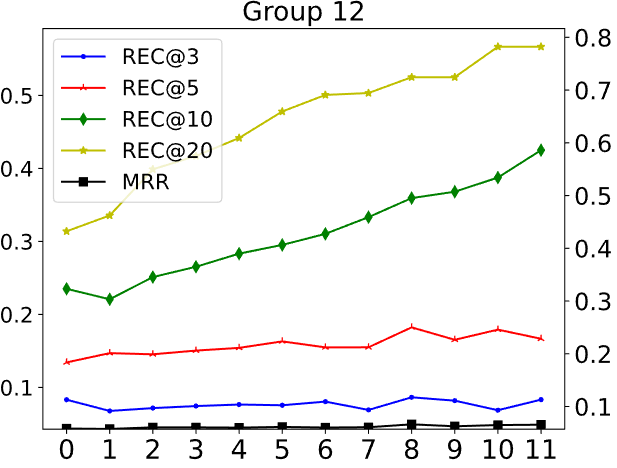}
\includegraphics[width=0.24\textwidth]{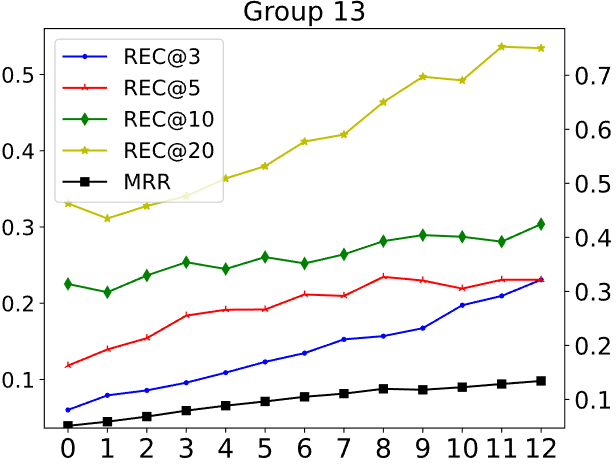}
\includegraphics[width=0.24\textwidth]
{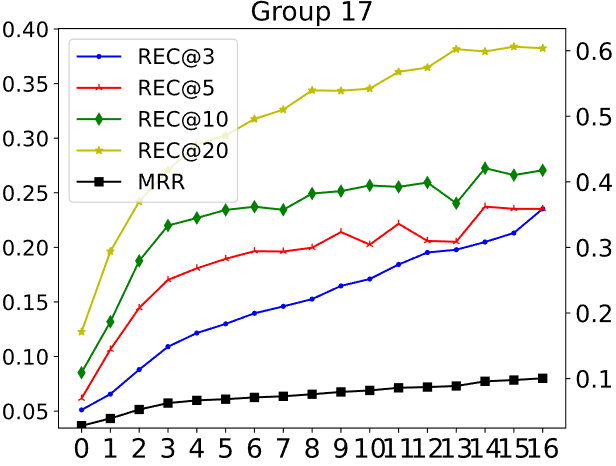}
\vspace{-10pt}
\caption{Recommendation result in terms of accuracy metrics in mashup creation process}
\vspace{-10pt}
\label{fig_metric_distribution}
\end{figure*}
Ideally, the recommendation of our approach shall help mashup development incrementally approaching mashup goal. In other words, as the creation step increases, the recommendation results of our approach should become increasingly better. In our test bed, we noticed that the number of creation steps (i.e., the number of composing services) differs in different mashups. Thus, to make it easier to answer \textbf{RQ2}, we divided the mashups into groups, each containing mashups with the same number of creation steps. For example, $g_{3}$ groups the mashups comprising three services. For a mashup $m_{j}$ in group $g_{l}$, we measured the recommendation performance over the number of creation step $t$ ranging from 0 to $l - 1$. At creation step $t$, the contextual services $S^{j}_{t}$ and ground truth services $G^{j}_{t}$ are generated as we described previously in Section \ref{sec_model_training}.

Fig. \ref{fig_metric_distribution} presents the average values of accuracy metrics over each number of creation step $t$ in different mashup groups $g_{3}{\sim}g_{17}$. 
For each sub-figure in Fig. \ref{fig_metric_distribution}, x-axis represents the index of creation step ($t$). The left y-axis and right y-axis are mean values of evaluation metrics of REC@K and MRR defined in Eqs.~\eqref{eq_17} and \eqref{eq_18}, respectively. The black line depicts the average values of MRR over all test cases (i.e., use the right y-axis). The other colored lines show the test results of REC@K when K=3,5,10,20, respectively. 

As shown in Fig. \ref{fig_metric_distribution}, for almost all testing groups, the recommendation performance goes up consistently as $t$ increases. In some groups, though, the accuracy metrics sometimes go down. 
The reason we found is that some services are only included in the mashups in the testing set. Thus, the embeddings of those services are never learned in the training phase. Examples are two out of eight services,  66,366\footnote{https://www.programmableweb.com/api/skimlinks-link-monetization} and 148,157\footnote{https://www.programmableweb.com/api/parsebot}, in the mashup 190,280\footnote{https://www.programmableweb.com/mashup/christmas-list-app} in group 8. 
Overall, considering the average ratio of such ``long-tail" services over all mashups is lower than 8.23\%, the experimental results align with our intuition: with the number of steps $t$ getting higher, the more information we could extract from the context, the better the recommendation performance could be, and the closer to mashup goal the recommended result is.

\vspace{-10pt}
\subsection{Effect of Attention Mechanism}
\begin{figure}
\vspace{-15pt}
	\begin{minipage}[t]{0.5\linewidth}
		\centering
  \includegraphics[width=\textwidth]{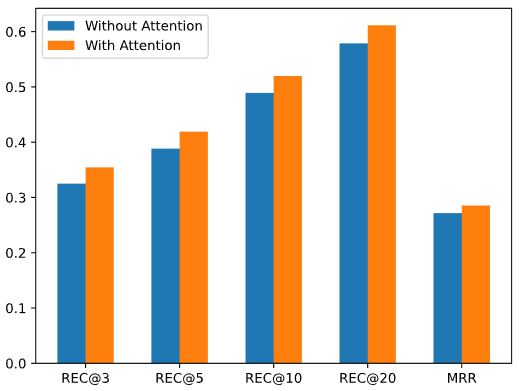}
		\vspace{-25pt}
		\caption{The effect of attention mechanism}
		\vspace{-25pt}
		\label{fig_attention_or_not}
	\end{minipage}
	\begin{minipage}[t]{0.5\linewidth}
		\centering
  \includegraphics[width=\textwidth]{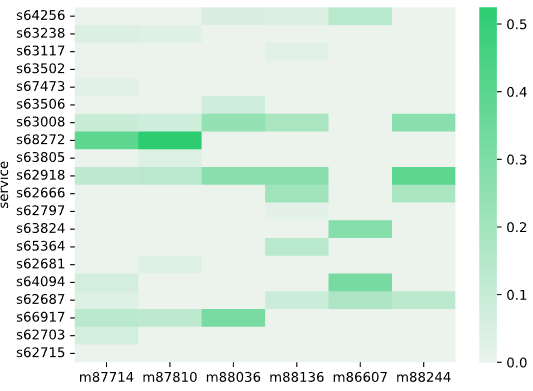}
		\vspace{-25pt}
		\caption{Visualization of attention weights}
		\vspace{-25pt}
		\label{fig_attention_heatmap}
	\end{minipage}
\vspace{-12pt}
\end{figure}



\vspace{-5pt}
In this subsection, we will answer the third research question \textbf{RQ3} by investigating the effect of the attention mechanism in our approach. 
Fig. \ref{fig_attention_or_not} shows the effect of the attention mechanism in terms of the accuracy metrics of REC@K and MRR. In terms of REC@3, REC@5, REC@10 and REC@20, the recommendation performance increases by 6.91\%, 6.69\%, 6.28\% and 5.71\%, respectively. 
In terms of MRR, we observe a similar result, as the performance increases by 3.79\%. The results are in line with the empirical fact that adopting the attention mechanism could enhance the expression ability of the embeddings of services and finally improve the overall performance.


Fig. \ref{fig_attention_heatmap} illustrates the visualization of attention weights on six example mashups and the corresponding 20 services. 
The horizontal and vertical coordinates are IDs of mashups and services in ProgrammableWeb, respectively. In mashup 87,810\footnote{https://www.programmableweb.com/mashup/sharemetric-chrome-extension}, two services contribute the most: a chrome extension pulling social network sharing counts into the chrome browser, and a service 68,272\footnote{https://www.programmableweb.com/api/pinterest-domain} for accessing detailed user data on Pinterest. 
This experiment shows that, given a set of selected services in the incremental mashup creation process, the attention mechanism is \emph{adaptive}, \emph{expressive} and \emph{context-aware} to represent the latent requirement of the unfinished mashup.

\vspace{-15pt}
\section{Conclusions \& Future Work}
\vspace{-12pt}
We regard mashup creation as an incremental process which at each step, we can help developers speed up the mashup development by recommending the next service. We have presented a novel machine learning method of goal-driven context embedding, which learns decision making strategies for service composition toward mashup goals in addition to service co-occurrence and semantic descriptions. To efficiently train parameters, we have introduced a goal exclusionary negative sampling strategy. Note that our approach is general and can be expanded to broader next candidate recommendation scenarios that are goal oriented, such as next product recommendation in e-shopping systems.

In the future, we plan to extend our research in the following three directions. First, we argue that a developer's preference plays an important role in her mashup creation process. 
We thus plan to take personalized information into account to further enhance our approach. Second, this work applies a na{\"\i}ve function, i.e., element-wise plus, for embedding fusion. We plan to explore other fusion functions to make our model more expressive. Third, in fact, in the incremental process of mashup creation, the order of contextual services is an influencing factor for next service selection. We will take the sequential creation behavior into consideration to further consolidate our approach.

\end{document}